\documentclass[twocolumn]{aastex61}
\usepackage{color}

\shorttitle{DEMONEXT}
\shortauthors{Villanueva Jr. et al.}

\begin{document}

\title{The DEdicated MONitor of EXotransits and Transients (DEMONEXT): System Overview and Year One Results from a Low-Cost Robotic Telescope for Follow-Up of Exoplanetary Transits and Transients}

\email{svillan@astronomy.ohio-state.edu}
\author{Steven Villanueva Jr.}
\affil{Department of Astronomy, The Ohio State University, 140 West 18th Av., Columbus, OH 43210, USA}

\author{B. Scott Gaudi}
\affil{Department of Astronomy, The Ohio State University, 140 West 18th Av., Columbus, OH 43210, USA}

\author{Richard W. Pogge}
\affil{Department of Astronomy, The Ohio State University, 140 West 18th Av., Columbus, OH 43210, USA}
\affil{Department of Astronomy and Center for Cosmology \& Astro-Particle Physics, The Ohio State University, Columbus, OH 43210, USA} 

\author{Jason D. Eastman}
\affil{Harvard-Smithsonian Center for Astrophysics, 60 Garden St., Cambridge, MA 02138, USA}

\author{Keivan G. Stassun}
\affil{Department of Physics \& Astronomy, Vanderbilt University, 6301 Stevenson Center, Nashville, TN 37235, USA}

\author{Mark Trueblood}
\affil{Winer Observatory, P.O. Box 797, Sonoita, Arizona 85637-0797, USA}

\author{Patricia Trueblood}
\affil{Winer Observatory, P.O. Box 797, Sonoita, Arizona 85637-0797, USA}

\submitjournal{the Publications of the Astronomical Society of the Pacific}

\begin{abstract}
We report on the design and first year of operations of the DEdicated MONitor of EXotransits and Transients (DEMONEXT). DEMONEXT is a 20 inch (0.5-m) robotic telescope using a PlaneWave CDK20 telescope on a Mathis instruments MI-750/1000 fork mount. DEMONEXT is equipped with a $2048\times2048$\,pixel Finger Lakes Instruments (FLI) detector, a 10-position filter wheel with an electronic focuser and $B$, $V$, $R$, $I$, $g'$, $r'$, $i'$, $z'$, and clear filters. DEMONEXT operates in a continuous observing mode and achieves 2--4\,mmag raw, unbinned, precision on bright $V<13$ targets with 20--120\,second exposures, and 1\,mmag precision achieved by binning on 5--6 minute timescales. DEMONEXT maintains sub-pixel ($<0.5$~pixels) target position stability on the CCD over 8 hours in good observing conditions, with degraded performance in poor weather ($<1$~pixel). DEMONEXT achieves 1--10\% photometry on single-epoch targets with $V<17$ in 5 minute exposures, with detection thresholds of $V\approx21$. The DEMONEXT automated software has produced 143 planetary candidate transit light curves for the KELT collaboration, and 48 supernovae and transient light curves for the ASAS-SN supernovae group in the first year of operation. DEMONEXT has also observed for a number of ancillary science projects including Galactic microlensing, active galactic nuclei, stellar variability, and stellar rotation.
\end{abstract}


\section{Introduction}

Time-series photometry is a useful tool for studies of exoplanets, supernovae, eclipsing binaries, and active galactic nuclei. Much of the night sky is dynamically changing, and synoptic observations can provide essential information. Despite the ever increasing demand for larger-aperture telescopes on the ground, as well as new and larger-aperture space-based telescopes, ground-based time-series photometry from small, half-meter aperture telescopes continues to be a powerful tool in modern astronomy. Many surveys continue to rely on small ground-based observatories as the workhorses of the follow-up observations and including facilities such as Las Cumbres Observatory \citep{brown13} and Minerva \citep{swift15}. In addition to their use to study known variable objects, such telescopes can play an important role as survey instruments, as demonstrated by a number of recent discoveries by surveys such as MEarth \citep{nutzman08} and Trappist \citep{gillon16}.

With the large number of targets to be discovered by upcoming all-sky surveys like TESS \citep{ricker15,sullivan15} and LSST \citep{ivezic08}, the demand for additional observations from these half-meter telescopes is unlikely to decrease in the near future. Due to the expected large numbers of transients from these surveys, a more efficient and effective follow-up tool is required, and suggests the need to develop and deploy more fully robotic telescopes. With this in mind, The Ohio State University and Vanderbilt University have come together to develop a new robotic observatory called the DEdicated MONitor of EXoplanets and Transients (DEMONEXT).

This paper is outlined as follows, in Section~\ref{secscience} we describe the scientific motivation for building the telescope. The optical and mechanical system is described in Section~\ref{secoptmech}. Nightly operations are described in Section~\ref{secoper}. The data products and system performance are described in Sections~\ref{secdata} and~\ref{secperform}. A summary of first year science results can be found in Section~\ref{results}, and conclusions in Section~\ref{secconcl}.

\section{Science Drivers}\label{secscience}

In order to maximize both the quantity and quality of data from DEMONEXT, we must first consider the science drivers for developing and deploying a relatively small-aperture, robotic telescope. There are two ongoing surveys that both The Ohio State University and Vanderbilt University are invested in, and what follows are the stated goals from the joint Ohio State and Vanderbilt community for DEMONEXT. These goals dictate the hardware selection, software design, and the operations model we have adopted.

\subsection{KELT Exoplanet Follow-Up}

The Kilodegree Extremely Little Telescope (KELT) ground-based transit survey \citep{pepper07,siverd12,pepper12,kuhn16} has reported sixteen exoplanet discoveries thus far\footnote{https://exoplanetarchive.ipac.caltech.edu/index.html}, including one joint discovery \citep{temple17}, and one simultaneous discovery \citep{lund17,talens17}, and has more planets in preparation (KELT Collaboration, private communication). KELT is primarily sensitive to hot Jupiters around bright host stars, typically $V<13$. KELT candidates' primary transits occur once per cycle with periods from days to weeks with the transit duration lasting 1--8 hours. Although the fraction of time for which one would want to make observations for any individual system is low, with close to 300 objects currently on the KELT North and South candidate lists, there is no shortage of candidates requiring observations each night. To constrain the planetary nature and to aide in future scheduling, it is crucial to observe the ingress or egress of each transit, as this confirms the depth and timing of future transits. Scheduling observations around hundreds of candidates only during transit becomes critical. Primary transit events for these massive planets have depths of $1\%$ or less, with millimagnitude (mmag) precision photometry required over the full duration of the event. Using relative aperture photometry, this can be achieved with 10--30 comparison stars in a single field-of view (FOV). During the transit event observations should be taken at the highest cadence possible, with exposure times of seconds to a few minutes, to constrain the shape, parameters, and timing of individual transit light curves.

To make these observations we require a telescope with sufficient aperture to make mmag precision follow-up observations on $V<13$ targets. This can be achieved with $<1$ meter telescopes. However, to avoid saturation, the telescope must be defocused prior to observations. To obtain a sufficient number of comparison stars using relative aperture photometry we require a $\sim0.5$\,degree diameter FOV. The telescope and mount should be stable, i.e. should achieve sub-pixel pointing, over the full duration of the observation, which may last over eight hours. This also requires continuous, uninterrupted observations, which excludes mounts that require meridian flips. The telescope must also have the ability to schedule around all of the 300 targets in a way that ensures that either an ingress or egress, preferably both, are observed for each transit, along with some amount of both out-of-transit and in-transit data.

\subsection{ASAS-SN Transient Follow-Up}

The All-Sky Automated Search for SuperNovae (ASAS-SN)\footnote{http://www.astronomy.ohio-state.edu/$\sim$assassin/index.shtml} surveys the entire visible sky every night down to $V=17$ \citep{shappee14}. Because ASAS-SN discovery images have large pixels, transient candidates require follow-up observations with lager telescopes and finer sampled images. Follow-up observations require a prompt single-epoch $V$-band image for candidate confirmation. It is important to confirm the transient nature as quickly as possible, in order to begin characterization as early as possible. Following confirmation, a month-long observing campaign is initiated to characterize the photometric evolution of the transient at the 1-10\% level. This campaign consists of multi-epoch, multi-band observations at 1--4 day cadence.

Follow-up of ASAS-SN targets requires a facility capable of both $1-10\%$ photometry in multiple bands at daily to weekly cadence, and a tool for prompt observations. For the faint ($V=17$) targets, $1-10\%$ photometry can be achieved in a single 5 minute exposure with a half-meter telescope. The guiding requirements are less strenuous. The pointing is required to drift less than a pixel over the exposure to ensure images are not elongated due to tracking errors. A filter wheel with a full compliment of either Sloan $g'$, $r'$, $i'$, and $z'$ or Johnson-Cousins $B$, $V$, $R$, and $I$ filters are required for multi-band characterization. The scheduler must be able to observe objects in a flexible manner, and ASAS-SN observations are a natural fit to fill in gaps between KELT observations. Additionally, prompt follow-up of recent events is preferred for targets that are identified in real time.

\subsection{Ancillary Science}

Given the strict scheduling requirements of KELT follow-up observations, and the slow-cadence required of ASAS-SN follow-up observations, there is a significant amount of free time left in the DEMONEXT queue to fill. With observations scheduled around high-cadence KELT or long-term monitoring of ASAS-SN targets, any science that can be done with either long or short cadence time-series photometry around $V<20$ sources are ideal targets for DEMONEXT. Within those parameters, there have been no shortage of suggestions from the DEMONEXT community on how to fill this time. Some of the more popular uses have been Galactic microlensing follow-up, monitoring of Active Galactic Nuclei (AGN), studies of stellar variability, stellar rotation, and eclipsing binaries. Many of these applications are described in Section~\ref{results}.

\subsection{Summary of Science Requirements}

In summary, DEMONEXT is required to make observations in two modes: a high-cadence, short-exposure time mode, around bright $V<13$ targets, and a low-cadence, long-exposure time mode around fainter $V<17$ targets. The telescope is required to automatically focus and defocus, depending on the target. For the low-cadence mode, guiding need only be stable enough to keep images from becoming elongated. However, for the high-cadence mode, in order to achieve $\sim$mmag photometry, sub-pixel stability is required over many hours with uninterrupted observations. The telescope must also be able to schedule around ingresses and egresses of roughly 300 KELT candidates, while filling in the gaps with observations of non-KELT targets. We also require a full set of either Sloan or Johnson-Cousins filters, with the ability to take multi-band observations. Finally, we need tools for prompt follow-up of rapidly-identified transient targets.

\section{Optical and Mechanical System}\label{secoptmech}

\begin{figure*}[t]
\centering
\includegraphics[width=16cm]{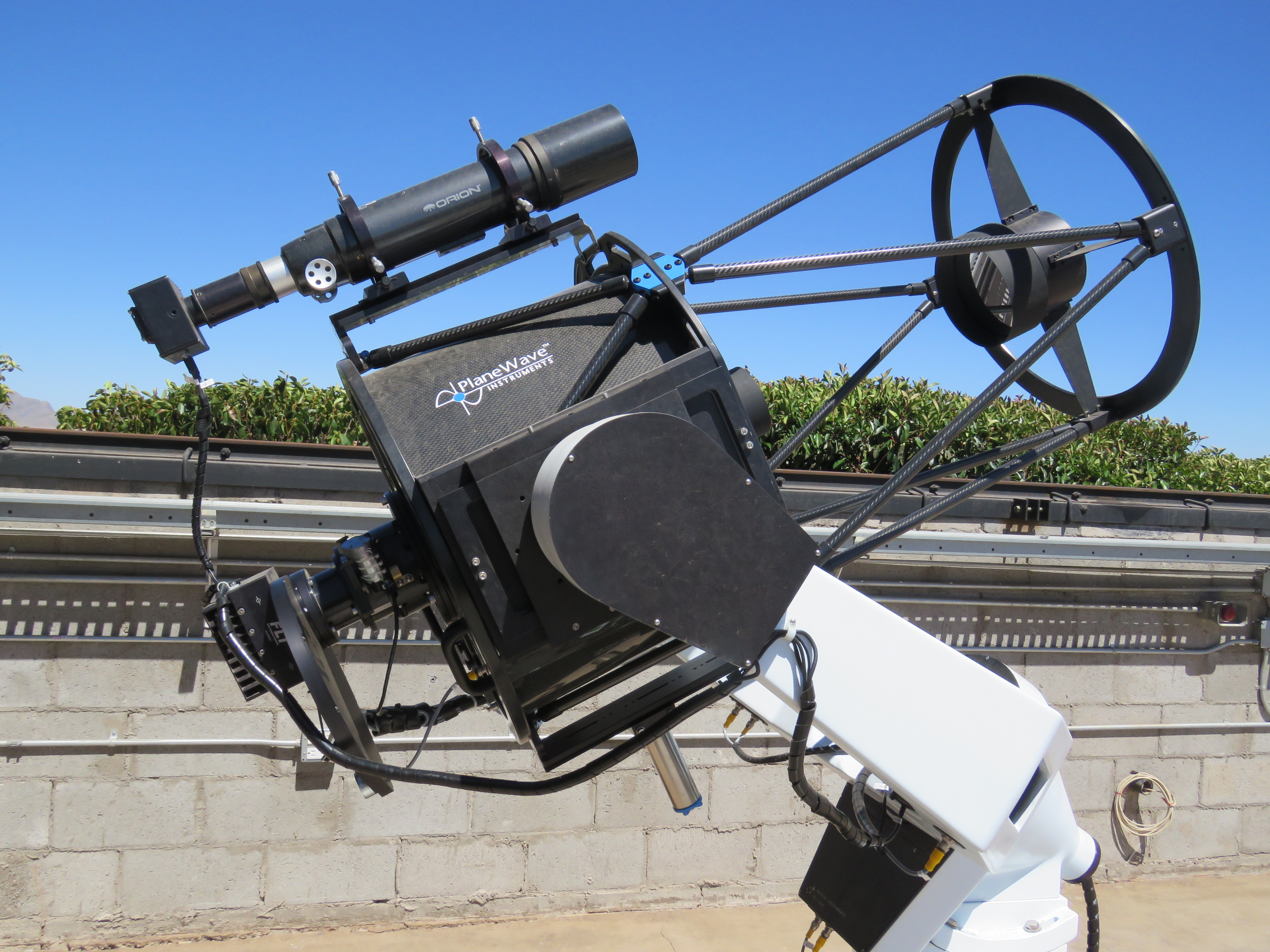}
\caption{\label{demonextscope}DEMONEXT is a PlaneWave CDK20 optical tube assembly on a Mathis Instruments 750/1000 mount pictured at Winer Observatory. DEMONEXT has a 2k$^{2}$ Finger Lakes Instrumentation CCD, a ten position filter wheel, and Hedrick electronic focuser. DEMONEXT has a SBIG guide camera and 80\,mm guide scope mounted above.}
\end{figure*}

The optical and mechanical details for DEMONEXT are largely unchanged from those reported in \cite{villanueva16} at the time of commissioning. We present a summary and description of the optical and mechanical system here, including updates made since commissioning in May 2016. A photograph of the final operational system is shown in Figure~\ref{demonextscope}, and a summary of the primary components is given in Table~\ref{tabsummary}. All components were purchased from vendors and integrated into one system by the authors. The three dominate costs were the telescope, mount, and science camera (30,000-45,000~USD each), and the entire system could be replicated for under 200,000~USD. Only a small number of parts, like mounting adapters or custom cables, were fabricated in-house.

\begin{table*}[t]
    \centering
    \begin{tabular}{c|c|c}
    \hline\hline
        Name & Model & Information\\
        \hline
        Telescope & PlaneWave Instruements CDK20 & 0.5\,m aperature\\
        Mount & Mathis Instruments MI-750/100 & equatorial fork\\
        Science Camera & Fairchild CCD3041 & 2k$\times$2k array, 0.9"\,pix$^{-1}$, 30.7$\times$30.7\,arcmin FOV\\
        Filter Wheel & Finger Lakes Instruments CFW-3-10 & Bessel(Johnson/Cousins) $B$, $V$, $R$, $I$, Sloan $g'$, $r'$, $i'$, $z'$, clear\\
        Focuser & PlaneWave Instruements Hedrick Focuser & 33\,mm travel distance\\
        Guide Telescope & Orion 80\,mm ED & 0.08\,m aperature\\
        Guide Camera & SBIG ST402-ME & 765$\times$510 array\\
        \hline\hline
    \end{tabular}
    \caption{Summary of the major parts and components that comprise DEMONEXT}
    \label{tabsummary}
\end{table*}

\subsection{Optical Telescope Assembly and Mount}

The DEMONEXT optical telescope assembly (OTA) is a PlaneWave Instruments 20\,inch (0.5\,meter) CDK20 telescope. The off-axis performance of the CDK20 allows for a field-of-view (FOV) greater than half-a-degree depending on the size of the detector array. The CDK20 telescope is mounted on a Mathis Instruments MI-750/1000 equatorial fork mount. This design allows for continuous and uninterrupted observations through the meridian, required for observations of long duration transit events or monitoring programs.

\subsection{Camera, Filter, and Focuser}
 
The science CCD and filters were re-purposed from DEMONEXT's predecessor, the DEdicated MONitor of EXotransits \citep{eastman10}. The imager is a 2k$\times$2k Fairchild CCD3041 detector packaged by Finger Lakes Instrumentation (FLI) with a 0.90\,arcsecond per pixel scale, and 30.7$\times$30.7 arcminute FOV. The pixel scale is well matched to the typical seeing of $\approx\,2$ arcseconds at Winer Observatory. DEMONEXT users are discouraged from requesting observations of high density fields or those requiring high spatial resolution, although users have had some success as shown in Section~\ref{secmu}. The large FOV results in an increased number of comparison stars available for differential photometry, the primary method of obtaining time-series photometry for the DEMONEXT community. DEMONEXT is equipped with a ten position FLI CFW-3-10 filter wheel that holds 8 filters: Sloan $g'$, $r'$, $i'$, $z'$, Bessell (Johnson/Cousins) $B$, $V$, $R$, $I$, and clear filters. We plan to add the final filter once a decision is made by the DEMONEXT user community. The CDK20 is equipped with a Hedrick electronic focuser, and PlaneWave;s electronic focus accessory (EFA) kit allows remote remote control of the telescope focus position with the PlaneWave Interface 3 (PWI3) software and drivers\footnote{http://planewave.com/downloads/software/}.

\subsection{Guide Scope}

A SBIG ST402-ME guide camera and an Orion 80\,mm ED refractor telescope were installed in October 2016 to auto-guide during continuous observations over many hours and for long ($>5\,$min) science exposures. The guide scope is aligned with the CDK20 and guides on the brightest star in the FOV. When used for KELT follow-up, the KELT target star is typically selected. For other observations, a guide star is selected within the FOV with guide camera integration times of $<10$\,seconds. The guide camera is not equipped with an automated focuser, and was focused manually during the installation. The focus does not change enough over time to require focusing more than once per year after summer shutdown restart.

\subsection{Remote Power and Connectivity}

Remote power management is implemented using an IP Power 9258 four outlet power supply. A single AC power cable is run to the power supply, which is connected to the science camera, guide camera, and mount. The IP Power 9258 allows us to individually power cycle each device via a local network connection. Each accessory's AC cable has an AC-to-DC converter. The main source of AC power for all telescope systems is filtered through a Powerware 9125 uninterruptable power supply. The focuser, filter wheel, and guide camera are powered using the accessory hub on the science camera.

The focuser, filter wheel, and guide camera are USB controlled through the accessory hub on the science camera, and the science camera, guide camera, and mount are USB controlled as well. To both minimize the number of cables that have to be run from the telescope to the control computer and overcome the 5 meter limit of most passive USB cables, we use an Icron 2304 four USB-to-Ethernet converter to convert the three USB lines to a single Cat 5 cable. We then convert that Cat 5 cable and IP Power 9258 Ethernet cable to a single fiber optic cable pair to transmit the devices over the 12.5 meters to the computer using a FiberTronics ESW-605 Ethernet-to-fiber switch. The only cables that run the full distance between the warm room and the telescope are a weatherproof AC power cable and an armored fiber optic cable pair. We convert the fiber optic back to Cat 5, and the Cat 5 to USB using the receivers of both switches.

\section{Nightly Operations}\label{secoper}

\subsection{Software}

The automation and scheduling software for the operational version of DEMONEXT are derived from those described in \cite{villanueva16}, but have been updated and are described in their current state here. A summary of the software and the components they interface with is given in Table~\ref{tabsoft}. All devices and software are Astronomy Common Object Model (ASCOM)\footnote{http://ascom-standards.org/} compatible, and come with a set of common scripted commands. An Optiplex 9020 desktop computer running Windows 7 has Python, MaximDL, PWI3, and Sidereal Technology software installed to execute all of the scheduling and software. All control programs were written by us in Python, using the Anaconda version of Python\footnote{https://www.continuum.io/anaconda} with the PyEphem\footnote{http://rhodesmill.org/pyephem/}, Numpy\footnote{http://www.numpy.org/}, astropy\footnote{http://www.astropy.org/}, and CV2\footnote{http://opencv.org/} modules. The Python programs written for DEMONEXT are based on older programs implemented in Visual Basic for the original DEMONEX system \citep{eastman10}, with new programs written specifically for DEMONEXT, or adapted from programs developed for the MINERVA Project\footnote{https://github.com/MinervaCollaboration/minerva-control} \citep{swift15}, which uses a similar mount to DEMONEXT. We also wrote a Python wrapper to initialize the auto-focus routine that comes with the EFA focus software. All of our Python code is available under the Creative Commons.

\begin{table}[t]
    \centering
    \begin{tabular}{c|c}
    \hline\hline
        Software & Instruments Controlled\\
        \hline
         & science camera\\
        MaximDL\footnote{http://diffractionlimited.com/product/maxim-dl/} & guide camera\\
         & filter wheel\\
        \hline 
        PWI3 & focuser\\
        \hline
        Sidereal Technology Servo II\footnote{http://planewave.com/downloads/software/} & mount\\
    \hline\hline
    \end{tabular}
    \caption{Summary of primary software purchased that operate DEMONEXT}
    \label{tabsoft}
\end{table}

\subsection{System Initialization}

During the afternoon of a scheduled night of operations, a master control program is executed to initialize the system. This program creates new engineering and observing logs for the night and performs various housekeeping tasks. This includes a query of the operations status file created at the end of each night to determine if DEMONEXT needs to execute hardware or software restarts to recover from fault conditions such as improper shutdown, power failure, etc. The night's transit and continuous monitoring programs are then scheduled as described in Section~\ref{sched}.

\subsection{Nightly Calibrations}

Each night we acquire calibration images before starting science observing. Bias and dark images are acquired before the roof is opened. Sky flats are acquired after the roof is opened in the evening when the Sun is $<3$\,degrees below the horizon. Because the time window for sky flats is so narrow, instead of attempting attempt to take flats in multiple filters during either morning or evening twilight, we take sky flats in the filter with the longest time elapsed since the last sky flat for that filter was taken. We scale the exposure times following each exposure to maintain counts in each sky exposure between 10--40k counts. On average, 14-17 flats with suitable counts are taken in two filters each night. This allows us to create a new master flat in each filter every 4--6 days depending on weather. Each master flat image has the standard deviation over the entire flat recorded along with the date. Standard deviations are 3--5\% across the illumination pattern. The best master flat used in calibrations each night is the master flat from all previous master flats with the lowest standard deviation. However, we add a constant term (0.01\%) to the standard deviation for each night elapsed since the observation to keep flats current.

\subsection{Observations}

The carbon fiber truss structure of the CDK20 makes for only small variations in focus with temperature. We focus the telescope twice per night: before the first science observations, and at the first convenient break between programs after midnight. We find the best focus in the $V$ band filter to within $\sim100$\,$\mu$m. Focus offsets relative to the $V$ band were measured in all other filters and the offsets range from 60-140\,$\mu$m. We note that these are all consistent with zero offset when considering the measured uncertainty for the best position each ($\sim100$\,$\mu$m). Nonetheless, we use the offset in each filter to re-position the focuser when changing the filter during observations. 

Science programs begin execution when the Sun is $>18$\,degrees below the horizon, and continue until either the Sun is $<18$\,degrees below the horizon or no science targets remain in the queue that can be completed before the end of astronomical twilight. DEMONEXT's operations are divided into two observing modes: objects requiring continuous observations (e.g. KELT Transit Targets) and the long-term monitoring queue (e.g. ASAS-SN Monitoring Programs), which require observations over many epochs. We do not yet utilize a mode for rapid follow-up as discussed in Section~\ref{asassn}. 

\subsubsection{Continuous Time-Series Observations}\label{sched}

We define our continuous time-series observations as a set of images taken in a continuous series, with the cadence set by the exposure time and overheads between exposures. We utilize this mode to take a series of a hundred to a thousand images consecutively on a single target over a span of one to eight hours. This can be done in a single filter or in an alternating sequence of filters. We use this mode for all of our observations of exoplanet transits from KELT or other target lists, monitoring of high priority Galactic micro-lensing events, and studies of stellar rotation.

DEMONEXT queries a list transiting exoplanet candidates, other targets requiring continuous observations, and a proprietary list of planetary candidates maintained by KELT known as the “K-list.” DEMONEXT calculates which targets' ephemerides indicate a transit will occur during the upcoming night while the target is visible. Viable candidates are sorted by priority, and the Julian Date (JD) of the start and end times for observations are scheduled. Excluding the times of observations of scheduled targets, any new targets that meet all of the requirements and can also be scheduled are included. The same process is repeated for the K-list, with the highest priority KELT targets scheduled first, to fill in as much of the night as possible. A minimum of 90 minutes is required for KELT observations and we require either the full ingress or egress. We avoid filling the queue with only in- or out-of-transit events, as neither provide interesting constraints on the depths or refine the ephemerides. We average 0--3 KELT candidates scheduled if there are no other targets in the queue.

Throughout the night, DEMONEXT queries the current time and references the start and end times of the scheduled transits. If the current JD is in the window of the scheduled transit or continuous observation start and end times, DEMONEXT will begin continuous observations. DEMONEXT will begin by slewing to the target's coordinates, moving to the best focus position(s) for the required filter(s), and sets the exposure time(s) if provided or calculates one according to a reference magnitude and a requested SNR. If multiple filters are requested, DEMONEXT alternates between each filter and sets the appropriate exposure time prior to each filter. Observations are repeated until the the scheduled end time. Including the 6.75 seconds of read time of the CCD, the total overhead between exposures is 12--18 seconds. This mode is used for covering transits or continuous-observation targets (e.g. observations of open clusters) for eight or more hours of uninterrupted imaging.

\subsubsection{Long-Term Monitoring}

We fill the times when no continuous observations are required by observing objects at low cadence in the long-term monitoring queue. The queue is populated by targets from a variety of sources, programs, and investigators and each target has a set priority, and a requested cadence. We use a dynamic scheduling system similar to the one used by the Remote Telescope System\footnote{http://rts2.org/} \citep{kubanek08,kubanek12}. Each object in the queue is given a score based on the current time, the time since the object was last observed, and the requested cadence of the observations. The score is modified by visibility constraints, priority, and partner time. The exact form of the scores are described in the Appendix of \cite{villanueva16}, although we modify these to meet our observational needs.

After calculating the score of each object in the queue, DEMONEXT observes the target with the highest score. We do not yet attempt to optimize the sequence of observations that would yield the highest set of scores, as there has yet to be any significant complaints on the quantity or quality of data from the investigators. For each observation, DEMONEXT will slew to the target's coordinates, re-position the focuser, set the filter wheel, calculate an exposure time, and acquire the requested exposures. The long-term monitoring queue is updated and the process is then repeated.

\section{Data Processing}\label{secdata}

All data processing is initiated manually at OSU using IDL\footnote{harrisgeospatial.com/ProductsandTechnology/Software/IDL.aspx} scripts each morning. Our IDL routines make use of the IDL Astronomy User's Library\footnote{https://idlastro.gsfc.nasa.gov/}. Calibrated images are available on the OSU servers for OSU users around noon each day, with delays over weekends. Vanderbilt images are pushed to a server at Vanderbilt following the reductions where they become available to the investigators. All images taken each night, including science images as well as bias, dark, and sky flats, are transferred to 10TB worth of storage at OSU, with a second 10TB hard drive for backup of all data. All data processing is done in IDL.

\subsection{Bias, Dark, and Flat Corrections}

Master bias and master dark frames are created by median combining all bias and dark images, respectively, which are used to bias and dark correct all images from that night. For the two filters of sky flats taken that night, a master sky flat is created by median combining the images in each filter. The master sky flats in every filter are stored in a separate directory. Each sky flat has the RMS recorded, and for each filter observed in that night, a single master sky flat in each filter is selected based on the RMS and time elapsed since the master sky flat was created. Each image is then flat corrected from the set of master sky flats chosen.

\subsection{World Coordinate Solution}

After bias, dark, and flat correcting the images, we execute IDL scripts to use the plate scale of the telescope, target RA and DEC, and FOV as inputs that are fed into SExtractor\footnote{https://www.astromatic.net/software/sextractor} to extract the locations of the sources in each image. We use the output source locations and Astrometry.net\footnote{https://www.http://astrometry.net/} to obtain World Coordinate Solution (WCS) on the calibrated images. The results are added to the fits header for each image. On nights with poor weather there are often too few stars to obtain a solution. Images without WCS solutions are typically discarded during the later analysis as a first order quality assurance check.

\subsection{Reductions Left to Users}

We provide users with science-calibrated images with WCS solutions. Given the number of DEMONEXT users and the variety of observations and science goals, we leave methods for analysis and time-series photometry up to individual DEMONEXT users. Each user is given the location on the OSU or Vanderbilt servers with their science calibrated images. Users are encouraged to copy the original images to their personal directories and perform their own aperture, PSF, or image subtractions routines.

\section{System Performance}\label{secperform}

\subsection{Noise}

\begin{figure}[t]
\centering
\includegraphics[width=8cm]{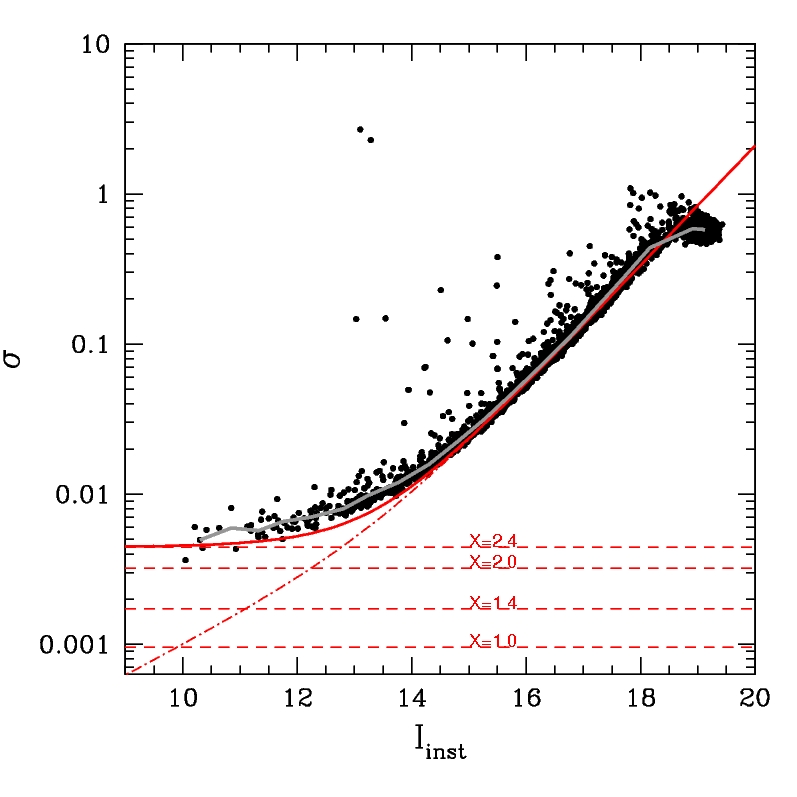}
\caption{\label{noise}Unbinned precision of DEMONEXT as a function of magnitude for the continuous observation mode. Diagonal dot-dashed line is the standard Poisson signal-to-noise expectation, while the scintillation noise floor at various airmasses are  shown as horizontal dotted lines. Observations were taken at airmasses from $X=1.5$--2.4 and the solid line is the $X=2.4$ and SNR lines added in quadrature. The gray line is the median in magnitude bins.}
\end{figure}

We investigate the noise performance of DEMONEXT photometry in the continuous time-series mode. The measured raw, unbinned, photometric noise as a function of magnitude is plotted in Figure~\ref{noise} from 8 hours of continuous observations of the Alpha Perseus cluster in $I$. The dot-dash diagonal line represents the standard signal-to-noise equation
\begin{eqnarray}
    SNR&=&\frac{S}{\sqrt{S+B+D+R^2}},
\end{eqnarray}
where we include terms for the source flux $S$, background $B$, dark current $D$, and read noise $R$. We do not include terms for flat fielding errors, comparison stars noise, fringing or other systematic effects. We include a term for the scintillation noise $N$ \citep{young67,hartman05}
\begin{eqnarray}
    N&=&N_{0}d^{-2/3}X^{7/4}e^{-h/8000}(2t_{exp})^{-1/2},
\end{eqnarray}
where $N_{0}=0.1$, the diameter of the telescope $d=50$\,cm, and the altitude of the observatory $h=1515.7$\,m are fixed. The exposure time for these images is $t_{exp}=20$\,s and the airmass varied from $X=1.5$--2.4 in these images. We plot a number of possible airmass terms in horizontal dashed lines. All lines, both SNR and scintillation, are calculated from the properties of the telescope and are not a fit to the data. We find that the data is well represented when the SNR and airmass $X=2.4$ lines are combined. The gray line is the median of various bins in magnitude to the DEMONEXT data. 

DEMONEXT regularly achieves 1\% photometry for targets with sufficient counts ($>10^{4}$) above the DEMONEXT airmass limit of $X=2.4$. When near zenith, DEMONEXT should achieve mmag photometry on sufficiently bright stars (i.e. counts $>10^{6}$), with 20 second exposures. In practice, DEMONEXT achieves 2--4\,mmag photometry on $V>12$ KELT targets over a range of airmasses and exposure times that correspond roughly to the scintillation noise. The scintillation noise floor is set by the photometric precision floor for most stars, but the bright end is dominated by a combination of scintillation noise, flat fielding errors, and fringing.

\begin{figure}[t]
\centering
\includegraphics[width=8cm]{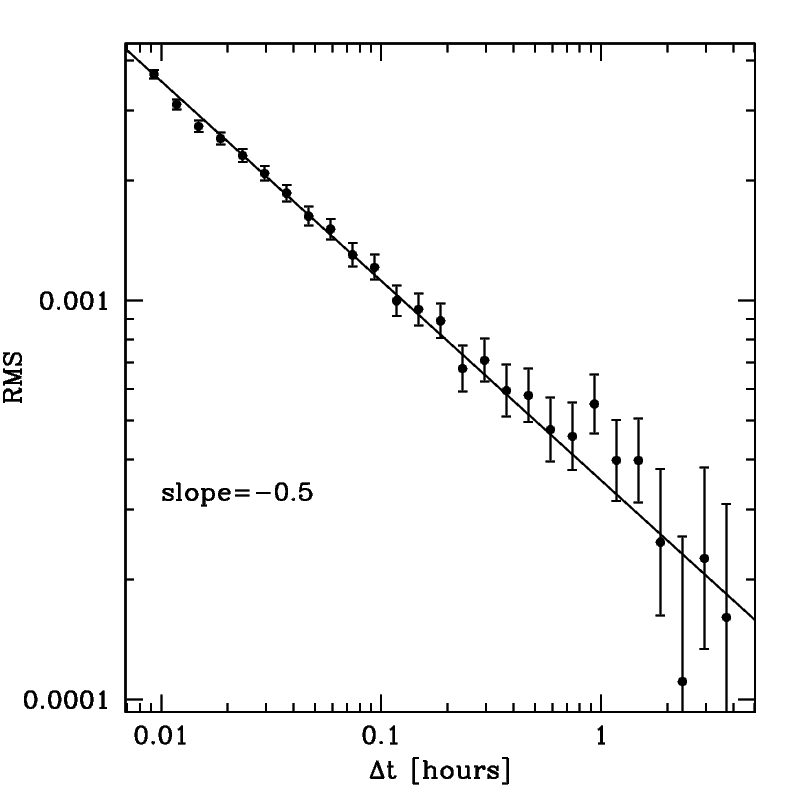}
\caption{\label{avlog}Allan variance plot of DEMONEXT data. The precision of the raw, unbinned data is the first data point in the top left. The precision of the binned data decreases as $\sqrt{t}$ as expected from white noise, but there is marginal evidence of red-noise on long timescales. The line is normalized to the first unbinned data point, but the slope is fixed.}
\end{figure}

We also verify that the noise decreases as $\sqrt{t}$ as expected of white noise. This can be seen in Figure~\ref{avlog}, where the raw, unbinned data of one of the brighter stars ($V\approx11$) is shown as the first data point in the top left. The binned data and their uncertainties are the data points. The predicted white-noise slope of $\sqrt{t}$ is shown in the solid line, normalized to the RMS of the unbinned data point. The data follow this slope, with only marginal evidence of non-white noise at longer time-scales. We therefore expect mmag precision at 5--6 minute timescales. 

\begin{figure}[t]
\centering
\includegraphics[width=8cm]{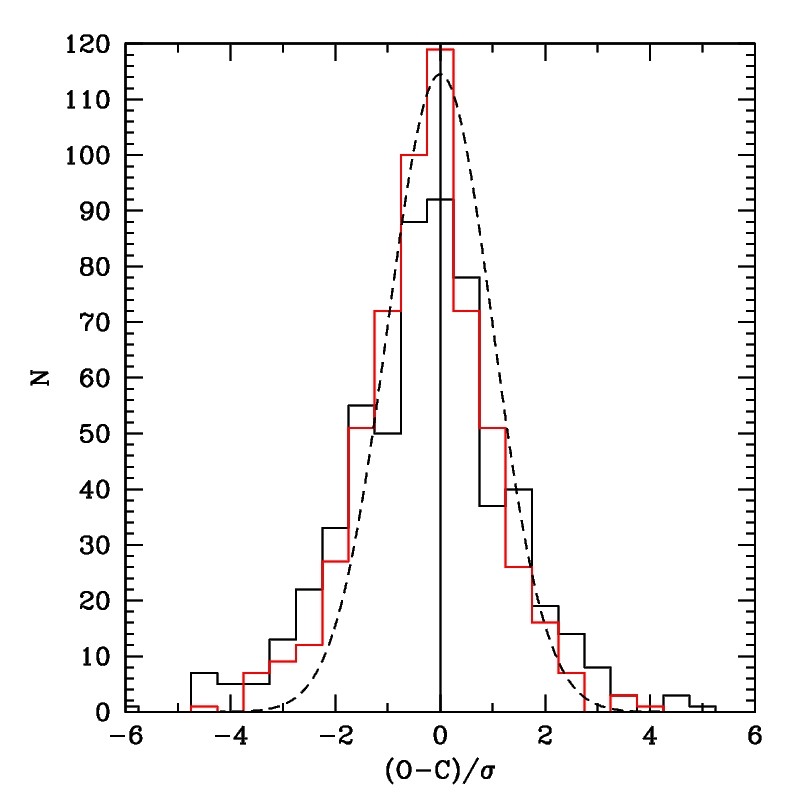}
\caption{\label{avh}Histogram of measured uncertainties (solid black histogram) compared to the expected uncertainty from photometric and scintillation errors (black dashed curve) with a $\chi^2/dof=2.89$. There are more data points with larger uncertainty than expected due to the systematic errors from flat fielding and fringing which we account for by scaling the errors to achieve a $\chi^2/dof=1.0$ (red histogram).}
\end{figure}

An initial look at the distribution of the errors shows that the errors are Gaussian with a $\chi^2/dof=2.89$ for a star with constant flux assuming photometric and scintillation noise. We model this data in Figure~\ref{avh} shown as the black histogram, and the expected distribution is shown as the black dashed curve. The difference in the actual distribution of errors (solid histogram) and the expected distributions (curve) is likely due to the systematic errors from flat fielding or fringing uncertainties. We scale the errors in the data to achieve a $\chi^2/dof=1.0$ as shown in the red histogram to account for the systematic uncertainties.

\subsection{Guiding}

DEMONEXT guiding errors can be described in two regimes: guiding errors incurred during a single exposure, and intra-exposure guiding errors that accumulate during continuous observations. For single exposures, guiding errors will cause individual point sources to be elongated. We have found that DEMONEXT drifts at a rate of up to $\le5\,\mathrm{pix\,hr}^{-1}$ ($\le4.5\,\mathrm{"\,hr}^{-1}$) when unguided. For a typical exposure, with a 5 minute integration time and a pixel scale of $0.9$\,"\,pix$^{-1}$, this causes the center of a point-source to drift by 0.083\,pixels (0.075\,"). With a typical seeing of 2\," or 2.22\,pixels, this results in an upper limit for the elongation of a point source within a single exposure of $\varepsilon\le0.036$ caused by guiding errors.

To obtain high quality, systematic-free light curves, DEMONEXT must minimize the effects flat-fielding errors, the loss of comparison stars that drift out of the FOV during observations, inter-pixel variations in sensitivity, and the effects of fringing (which DEMONEXT suffers from in the $I$ and $z'$ bands), by keeping targets on the same pixel for observations as long as 8 hours or more. Prior to the installation of the guide scope, DEMONEXT was used without any guiding. With a drift rate of $5\,\mathrm{pix\,hr}^{-1}$, drifts of 40\,pixels (36") were found over the course of 8 hours. After the installation of the guide scope, the drift was reduced to to $3\,\mathrm{pix\,hr}^{-1}$ ($2.7\,\mathrm{"\,hr}^{-1}$). Drifts of 30\,pixels (27") were still found in the longest observations, consistent with differential flexure between the guide scope and the OTA. The camera is mounted on the back of the OTA (see Fig.~\ref{demonextscope}), and a modification to the guide scope's mount is planned for Fall 2017 to increase the rigidity and stability of the system. To meet our sub-pixel guiding requirement, the modified guide scope mount would have to reduce the drift rate to $\le0.1\,\mathrm{pix\,hr}^{-1}$ ($\le0.09\,\mathrm{"\,hr}^{-1}$), an order of magnitude lower than the current rate. We investigate alternative methods to overcome the mechanical limitation of the guide scope.

\subsubsection{Science Guiding}\label{sciguide}

To meet the sub-pixel guiding stability DEMONEXT switched to guiding on the science images during spring 2017. By guiding on the science images themselves, we bypass the guide scope and any errors due to differential flexure. Prior to twilight, DEMONEXT takes a series of five images. The first is a reference, followed by two images after moving the mount twice in only RA, $\pm\alpha$, and then two more images after moving twice only in DEC, $\pm\delta$. DEMONEXT then solves for the locations of the brightest 50 objects using routines in the Python OpenCV library. For the stars in the FOV of each image, we measure $\Delta x$ and $\Delta y$ in the image plane and then fit for image rotation to map $\Delta\alpha$ and $\Delta\delta$ in RA and DEC, the relative rotation $\theta$, and scale factors $C_{1},C_{2}$:
\begin{eqnarray}
        \Delta\alpha&=&C_{1}\Delta x\cos{\theta}-C_{2}\Delta y\sin{\theta}\\
        \Delta\delta&=&C_{1}\Delta x\sin{\theta}+C_{2}\Delta y\cos{\theta}
\end{eqnarray}
The orientation and scale do not change from night-to-night, but are required to be measured as the CCD is not perfectly aligned North and East. We find that the CCD is mis-aligned with North up and East left to by $\approx$6\,degrees.

During continuous observations (e.g. transits) the first image is used as a reference image, with the locations of the 50 brightest sources recorded. Following each exposure (typically 20 seconds), DEMONEXT solves for the locations of the 50 brightest sources, and we solve for the $\Delta x$ and $\Delta y$ in the image plane to send a correction in RA and DEC ($\Delta\alpha$,$\Delta\delta$) to the mount prior to starting the next exposure. The total overhead is 12--18 seconds, and the additional overhead penalty due to science guiding is $<2$\,seconds, which is dominated by the time to execute the two slews. As the slew times are short $<1$\,second each, we are currently investigating ways to further reduce overheads.

\begin{figure}[t]
\centering
\includegraphics[width=8cm]{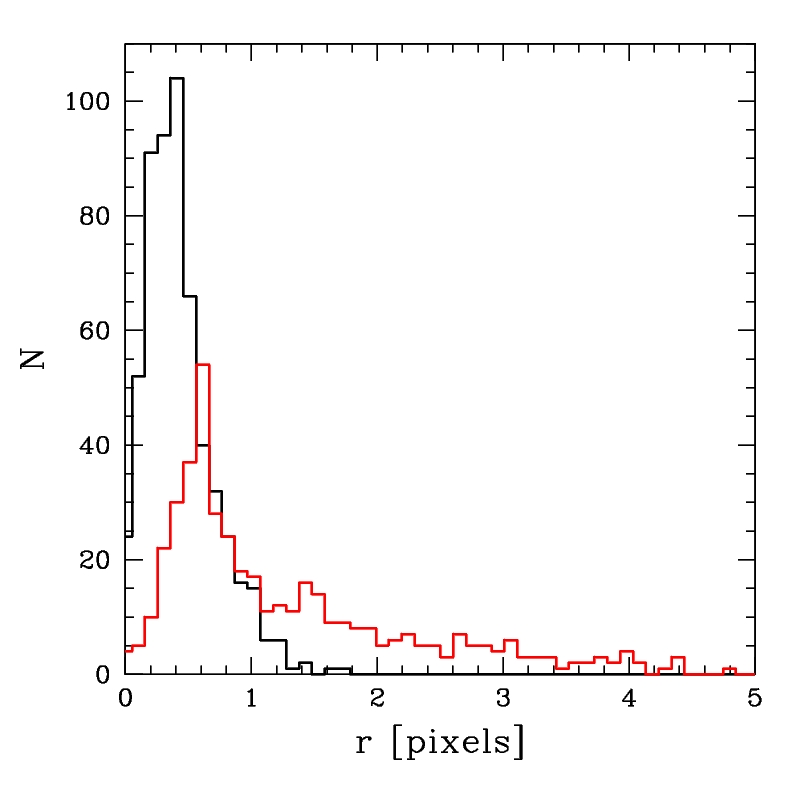}
\caption{\label{guiding}Histograms of the number of images with the positions of target stars relative to the target stars' mean for two representative nights. The black (red) curve is a night with good (bad) weather and observing conditions. For the night with good (bad) conditions, there are $N=576$ ($N=443$) images spanning 5.5 (4.4) hours. The target is kept stable with a median deviation of 0.44 (0.95) pixels.}
\end{figure}

We show two representative nights of science guiding in Figure~\ref{guiding}. The black line represent a night with good weather and no clouds. There are $N=576$ consecutive 20 second exposures (plus 12--18 seconds overhead), where target star stability is kept on the same pixel with the median deviation of 0.44 pixels and the peak in the distribution of 0.41 pixels. The red curve represents a night with poor observing conditions, with a constant layer of clouds. There are $N=443$, 20 second images with a peak in the distribution of 0.61 pixels, but with a longer tail and median deviation of 0.95 pixels. We find no evidence of long-term drift when using science guiding over 5.5 and 4.4 hours of guiding respectively. We are able to keep the median deviation to $<1$\,pixel in poor observing conditions, and $<0.5$\,pixels in good observing conditions. This is a substantial improvement over either not guiding, or using the guide scope. We have not yet investigated the possibility of using both the guide scope and science guiding in parallel to further increase guiding stability. With science guiding, DEMONEXT now meets the sub-pixel guiding requirements set forth by our science drivers.

\subsection{Automation Yields}

\begin{figure}[t]
\centering
\includegraphics[width=8cm]{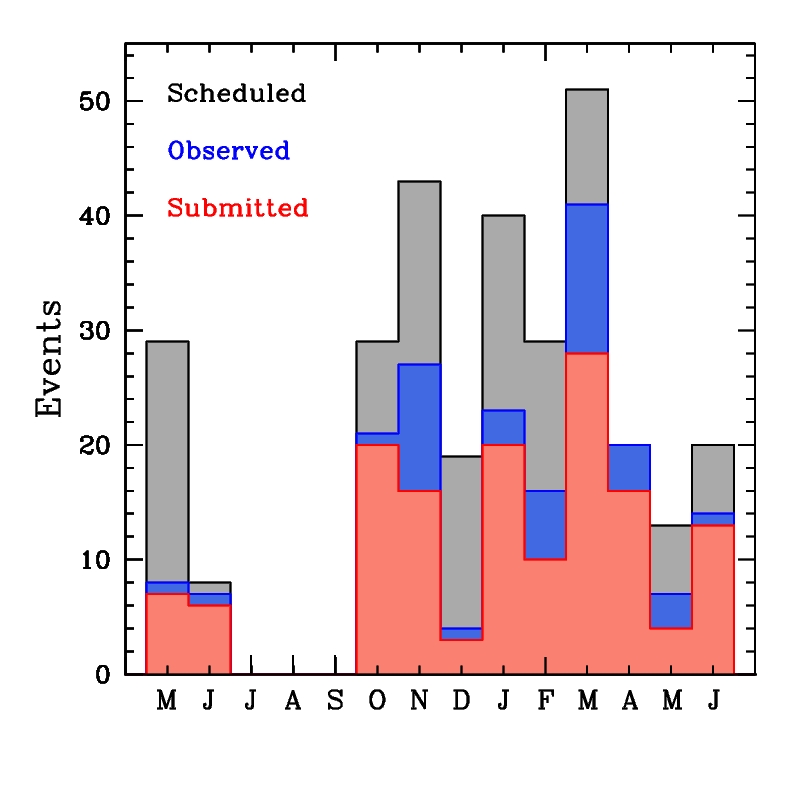}
\caption{\label{auto}Histograms of the number of DEMONEXT KELT transit events from 291 nights from May 2016 through June 2017. The gray histogram is the number of transit events scheduled per month, blue were those that were observed, and red are those submitted to the KELT collaboration. Features such as summer shutdown (July--September 2016) are visible. Discrepancies between the black and blue histograms can be attributed to commissioning difficulties (May 2016), weather, or other technical difficulties. Discrepancies between blue and red histograms are attributed to poor weather, poor data quality, or the failure of the filter wheel (March 2017). 76\% of observed light curves observed were of sufficient quality to submit to the collaboration, and DEMONEXT averaged 14--15 light curves per month during the first 291 nights of operations.}
\end{figure}

One of the primary reasons for automating the DEMONEXT telescope was to increase the yield of observations of KELT transiting planet candidates. To date DEMONEXT has been operational for 291 nights and submitted 143 transit light curves, including multiple events on the same night, or the same event on one night but in alternating filters. This is a rate of 14.7 events per month, or 0.49 per night. \cite{villanueva16} predicted a rate of 0.81 per night, but failed to include weather or ancillary science projects. Adjusting the \cite{villanueva16} rate for weather gives a predicted rate of 0.65 per night, comparable to the rate 0.49 per night we achieved. Distributions of scheduled, observed, and submitted events from May 2016 through June 2017 can be seen in Figure~\ref{auto}.

Differences between the number of scheduled events (black) and observed events (blue) can be attributed to a combination of technical issues and weather. Differences between the number of observed events (blue) and those submitted (red) are due to the data taken being unusable. Winer observatory closes during the summer monsoon season and accounts for the lack of observations in June-September 2016. The lack of usable observations can be seen during the month of May 2016 while in the early days of commissioning and general debugging, during November 2016 through March 2017 when we had a run of poor weather resulting in occasional roof closures, and we were commissioning a number of new features during the first year that all had problems to be solved. Small numbers of scheduled events can be seen during months cut short by the monsoon season (June 2016), during months with short nights (April--June 2017), and during months where other science projects were observed at a greater priority (December 2016). Cases where the data was unusable are usually due to weather, pointing or slew issues, and the failure of the original filter wheel (March 2017).

Overall DEMONEXT is performing well as a follow-up resource. We scheduled 301 transits in the first 291 nights of operation, averaging 31 events scheduled per month, excluding summer shutdown. Of those 301, 188 were observed (62\%) and 143 were submitted (48\%). Of those that were observed, 143 of 188 (76\%) were of sufficient quality to submit to the collaboration. Over the first year of operation, DEMONEXT yields an average of 14--15 usable light curves per month. We expect this number could increase to 24 per month assuming 80\% of nights are usable due to weather, zero mechanical or software malfunctions, and no other higher priority science programs.

\section{DEMONEXT Science Results}\label{results}

\subsection{KELT Exoplanet Follow-Up}

DEMONEXT observed 188 planetary transit events around 116 unique targets and submitted 143 transit candidate events to KELT. Targets are transiting planet candidates, and DEMONEXT is used to vet candidates for astrophysical false positives e.g., nearby eclipsing binaries. KELT candidate host stars have magnitudes that range from $7<V<13$ with planets that produce transit depths of 1--10\,mmag. Typical observations include a continuous series of exposures in a single band, most commonly $i'$ to minimize the effects of limb darkening. DEMONEXT also has the capability to observe in multiple filters to obtain wavelength-dependent transit depth measurements. For each target an exposure times is calculated to produce a SNR of $\sim$1000, which corresponds to exposure times of 120 seconds or less for most KELT stars. For objects that have exposure times less than 20 seconds, DEMONEXT adopts a minimum exposure time of 20 seconds and defocuses the telescope to spread the flux over more pixels to avoid non-linearity effects and saturation in the CCD. This is done to prevent the readout overhead from exceeding the shutter open time.

\begin{figure}[t]
\centering
\includegraphics[width=8cm]{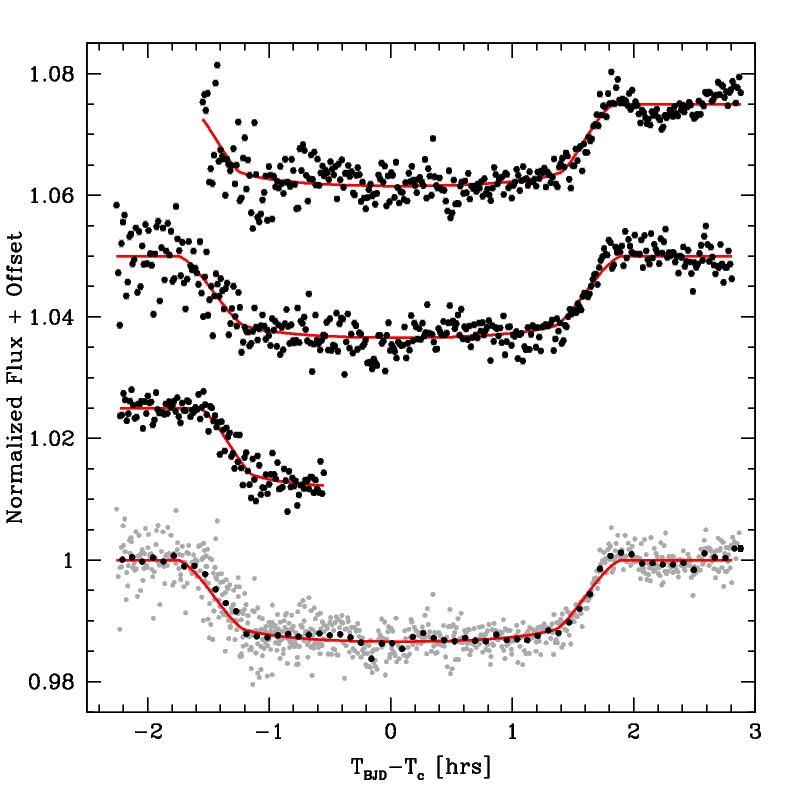}
\caption{\label{keltlc}DEMONEXT observations of KELT-20b/MASCARA-2b \citep{lund17,temple17}, the first planet confirmed with DEMONEXT data. DEMONEXT observed three transits from May--June 2016 of the V=7.58 host star with 31 second exposures. The top three light curves (black points) are detrended with respect to the target's position on the chip and airmass. The light curves have 2--4\,mmag RMS, depending on the airmass, relative to the best fit model (red curve). The bottom light curve has all three DEMONEXT light curves phased (gray points). Over-plotted is the phased and binned DEMONEXT light curve (black points) that has 0.89\,mmag RMS relative to the model (red curve).}
\end{figure}

The DEMONEXT light curves for KELT-20b were featured in \cite{lund17}, and we show these light curves in Figure~\ref{keltlc}. We note that this planet was independently discovered as MASCARA-2b \citep{talens17}. The three light curves shown were some of the earliest DEMONEXT light curves, but are the first for a confirmed exoplanet. The host star has $V=7.58$ and the transit depth of $\Delta V\approx14$\,mmag is clearly detected in the DEMONEXT data. Also visible are the effects of scintillation noise causing photometric accuracy to deteriorate at increasing airmass as the top two transit observations in Figure~\ref{keltlc} started at high airmass and the point-to-point scatter improves at later observations at lower airmass. These three light curves were taken in May and June 2016 during early commissioning, and are representative of the quality of the earliest DEMONEXT light curves. Since the implementation of science guiding (\S~\ref{sciguide}), DEMONEXT routinely achieves 2--4\,mmag precision on KELT targets, with mmag precision on data binned on 5 minute timescales.

\subsection{ASAS-SN Monitoring and Confirmation}\label{asassn}

\begin{figure}[t]
\centering
\includegraphics[width=8cm]{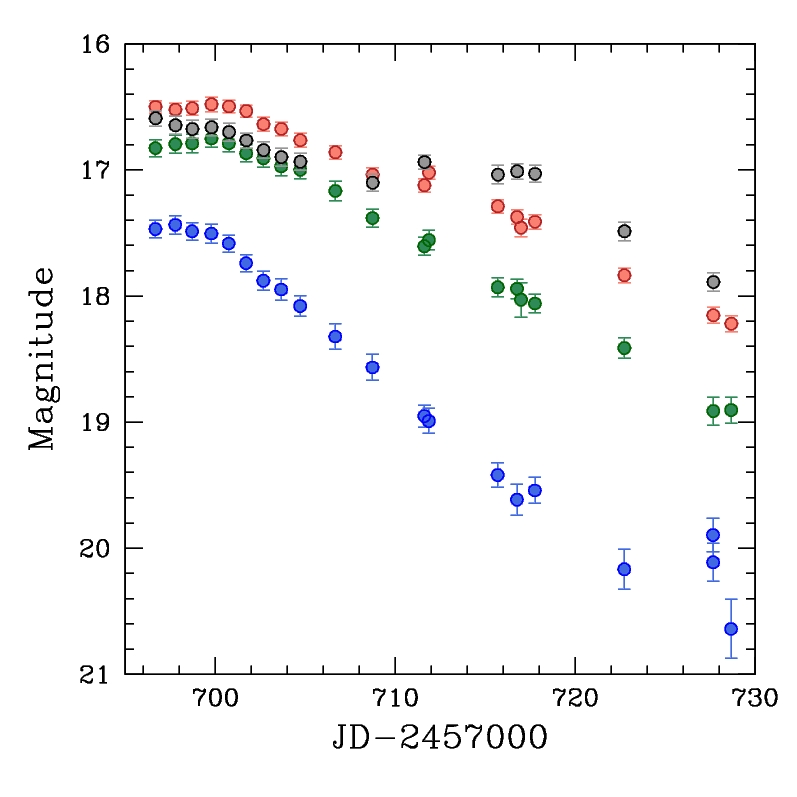}
\caption{\label{asassnlc}Light curve of type Ia supernoava 2016hli observed by DEMONEXT in 20 epochs. Requested cadence was once every 2 days in $B$, $V$, $R$, and $I$ filters (blue, green, black, and red points). Visible in the DEMONEXT data is the secondary peak for Type Ia in redder bands ($R$, $I$).}
\end{figure}

DEMONEXT has observed 48 ASAS-SN targets to date. Our primary follow-up targets are candidates for supernovae or tidal disruptions events with magnitudes that range from $V=15$ at peak and can fade to $V\sim21$ before being removed from the DEMONEXT queue. Typical observations include a sequence of four images in $B$,$V$,$R$,$I$ with exposure times from 180--300 seconds depending on magnitude. Objects remain in the DEMONEXT queue for around a month, and are observed at a typical cadence of once every 2 nights, although the cadence varies depending on the object.

As an example, Figure~\ref{asassnlc} shows the light curve of SN2016hli, a low-luminosity type Ia supernovae at redshift $z=0.017$. SN2016hli was observed by DEMONEXT in B,V,R,I in 20 epochs over 33 nights. The requested cadence was once every 2 nights, but the actual cadence of observations varies, from multiple observations in a single night to five day gaps, depending on other objects in the queue, demand from KELT targets, and weather. Visible in the DEMONEXT data is the secondary peak for Type Ia in red bands ($R$,$I$), that are seen in all but the lowest luminosity Ia. DEMONEXT was still able to detect 2016hli even at $V\approx20.5$.

\begin{figure}[t]
\centering
\includegraphics[width=8cm]{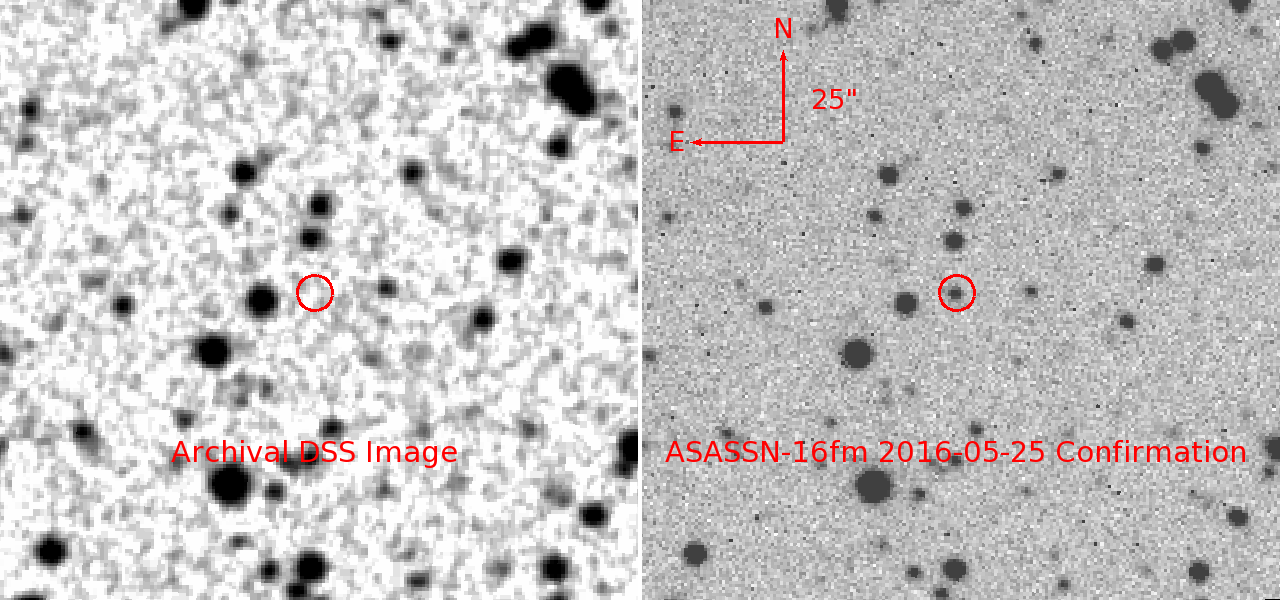}
\caption{\label{asassnpic}Confirmation of ASASSN-16fm taken by DEMONEXT featured in \cite{villanueva16b}. DEMONEXT images (right) are compared to archival SDSS images (left) to verify the presence of a transient objects.}
\end{figure}

Attempts were made to confirm targets identified by ASAS-SN in real time using a prompt follow-up tool. In most cases a single-epoch single-band "raw" image is all that is required to confirm a target for monitoring. Most follow-up observation requests are submitted within a few hours of discovery, and only the initial confirmation image is used. DEMONEXT images are not science ready until noon the next day, so DEMONEXT was unable to submit images before other observatories. A such, we have decided to disable this fast-confirmation observing mode until we can develop a fast reduction system. A serendipitous exception occurred when the author was troubleshooting the telescope during the call for the the confirmation of ASASSN-16fm. We were able to observe and submit a "raw" image prior to the rest of the community \citep{villanueva16b}, demonstrating the utility of rapid confirmation images if we can address the fast reduction issue.

\subsection{Microlensing Surveys and Follow-Up}\label{secmu}

\begin{figure*}[t]
\centering
\includegraphics[width=16cm]{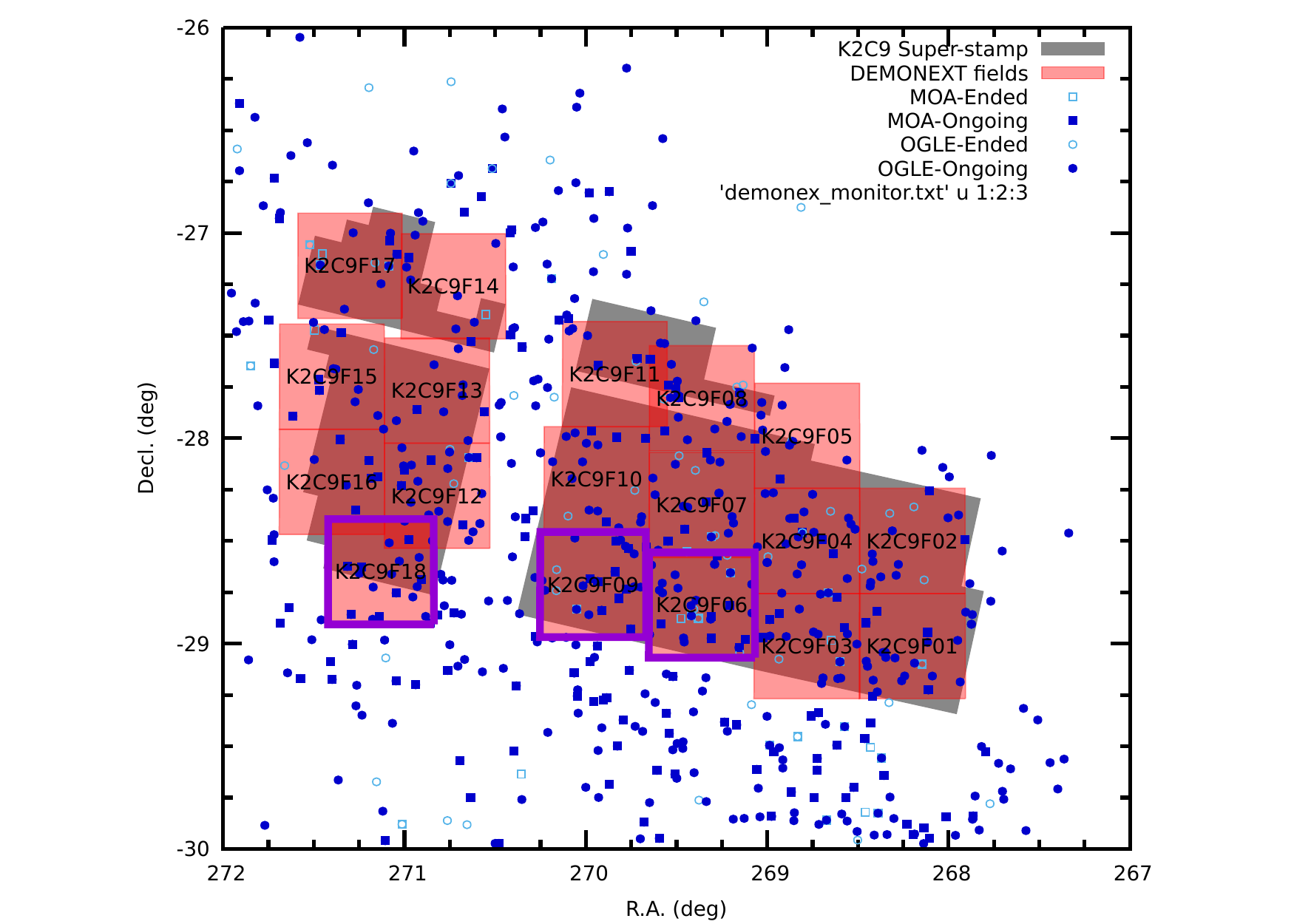}
\caption{\label{ulensingfields}DEMONEXT tiled the K2 Campaign 9 field in a microlensing campaign to simultaneously monitor the bulge from both the ground and space. DEMONEXT tiled the super-stamp (grey region) with all 18 fields (red areas) in 2 hours each night. Initially reduced fields are outlined in purple, while events detected in other surveys are shown in shades of blue.}
\end{figure*}

\begin{figure}[t]
\centering
\includegraphics[width=8cm]{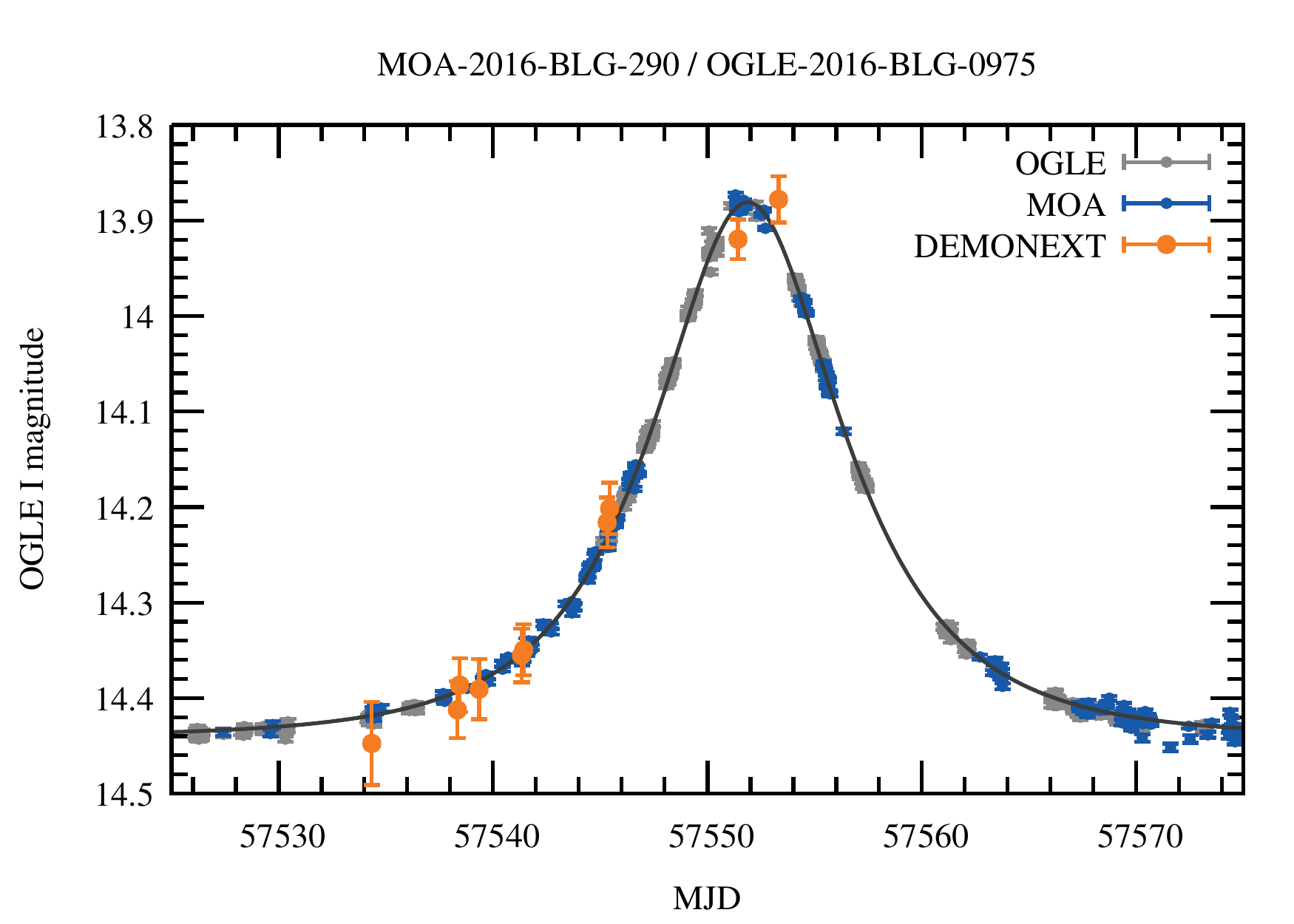}
\includegraphics[width=8cm]{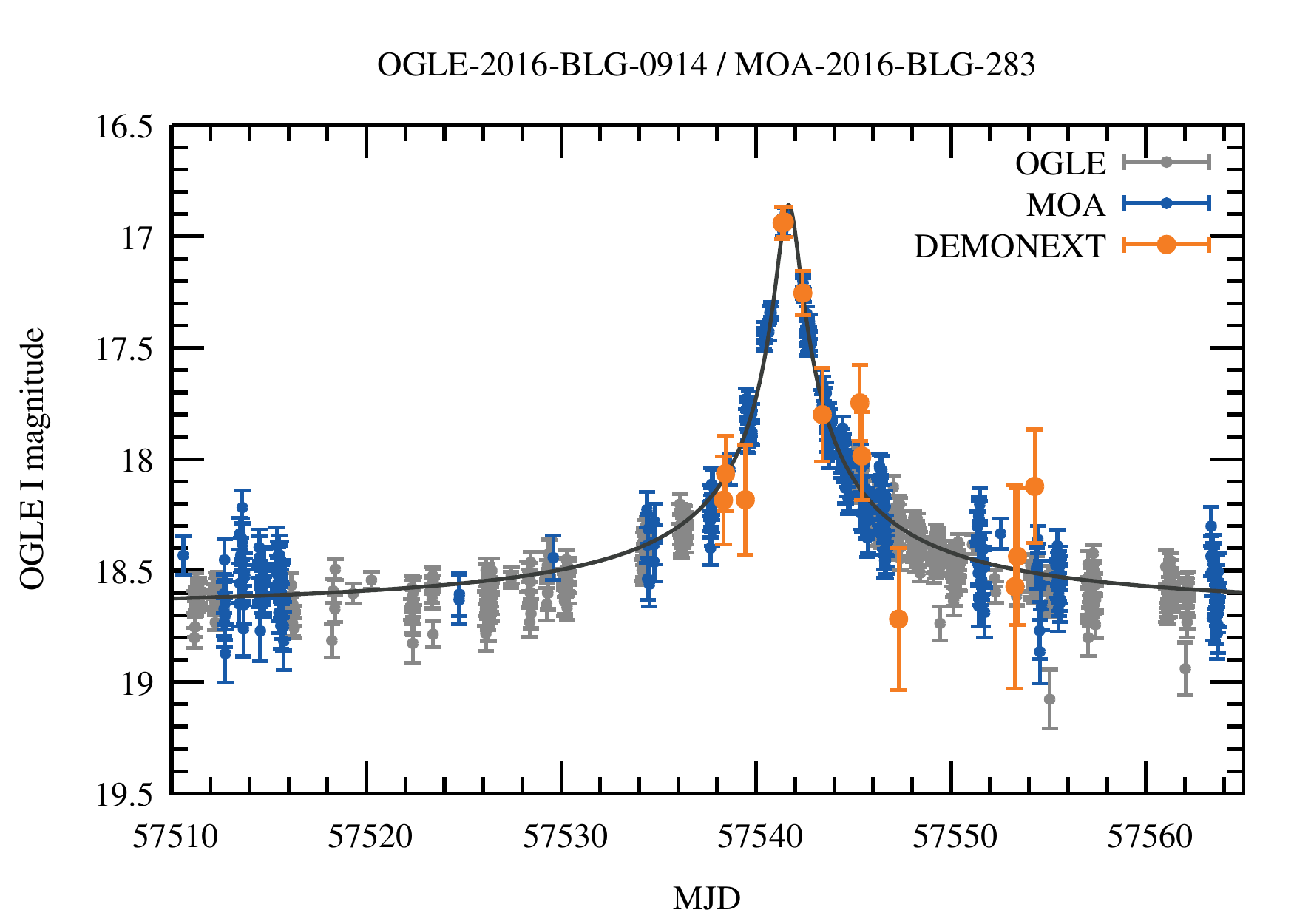}
\includegraphics[width=8cm]{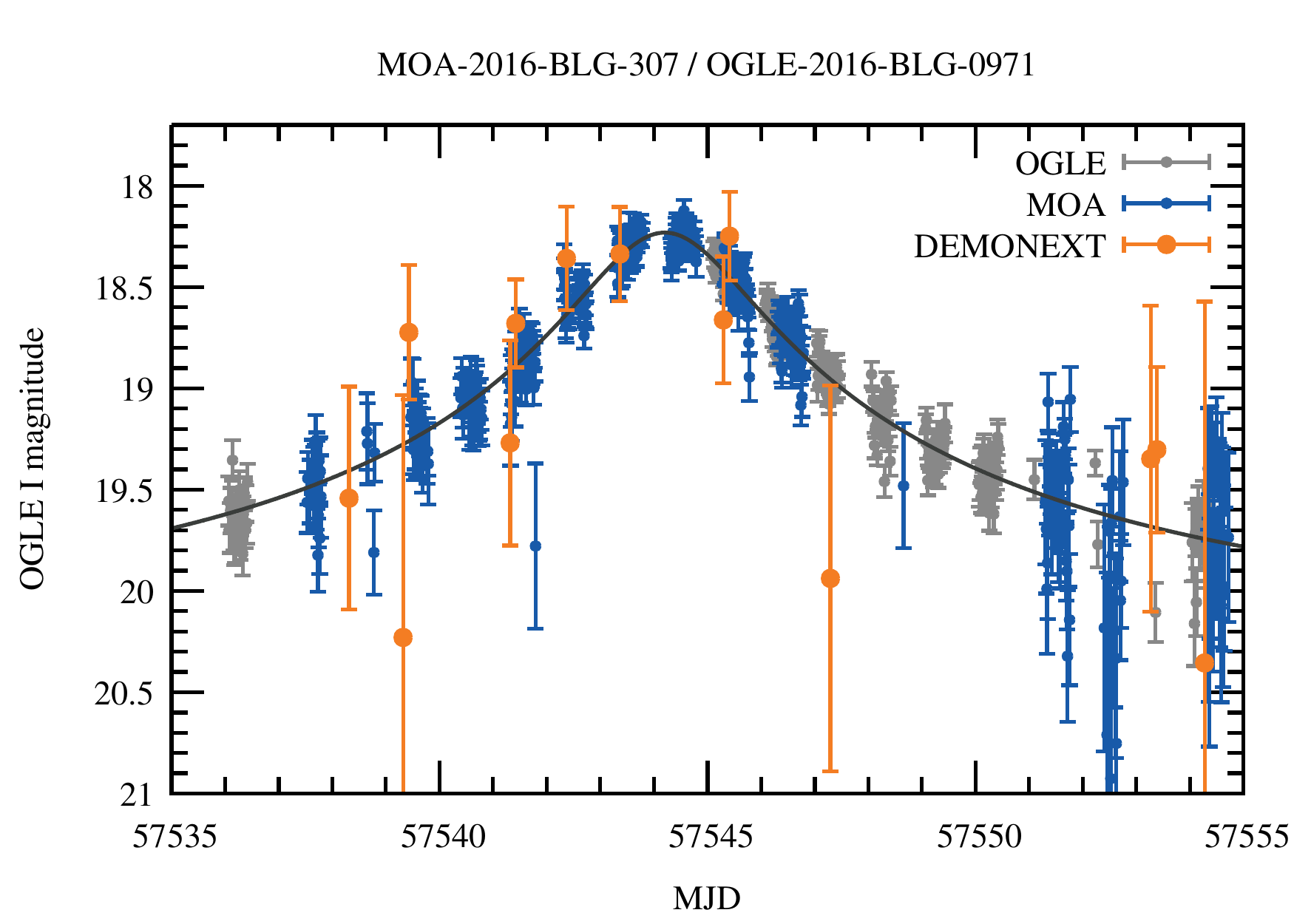}
\caption{\label{ulensing}Three microlensing events observed by DEMONEXT shown from a robust detection in the top panel, to a marginal detection at the bottom panel. Data from other surveys are in blue and grey, while DEMONEXT data is shown in orange and at lower cadence.}
\end{figure}

In addition to the planned observations of KELT and ASAS-SN targets, DEMONEXT has observed a number of ancillary science programs. Most notable is the observation of microlensing events. DEMONEXT microlensing targets are typically identified by the OSU users from other surveys such as OGLE \citep{udalski03}, MOA \citep{sako08}, and KMTNet. The requested observations are of $I<20$ objects in low cadence mode at daily cadence to establish baseline observations or to cover the other surveys during times of poor weather. High-cadence mode is used during caustic crossings or other perturbations of interest. In both cases, DEMONEXT has been used to provide data when the other surveys would saturate, typically at $V<13$.

DEMONEXT can also be used as a survey telescope. During May and June 2016, DEMONEXT participated in the K2 Campaign 9 microlensing campaign to simultaneously monitor the Galactic Bulge from both the ground and space. DEMONEXT tiled the K2 Campaign 9 super-stamp with a 1 day cadence of 18 fields in the $I$ band. In the low-cadence mode, DEMONEXT was able to observe all 18 fields in 2 hours, although the airmass limit was increased from 2.4 to 2.7 to do this. The large 0.5 degree FOV allowed us to tile the entire super-stamp as shown in Figure~\ref{ulensingfields}. From the reduction of the DEMONEXT Field 18, three microlensing events were observed by DEMONEXT are shown in Figure~\ref{ulensing}. We select these three events to show a brighter event at the top, while progressing to the detection limit of DEMONEXT at the bottom. Although DEMONEXT does not have a sufficiently high cadence to recover planetary microlensing signatures, it is able to recover microlensing events, even around sources as faint as $I\approx20$, in order to provide robust baseline coverage of these events, which is crucial for modeling any microlensing anomalies.

\subsection{Open Clusters}

\begin{figure}[t]
\centering
\includegraphics[width=8cm]{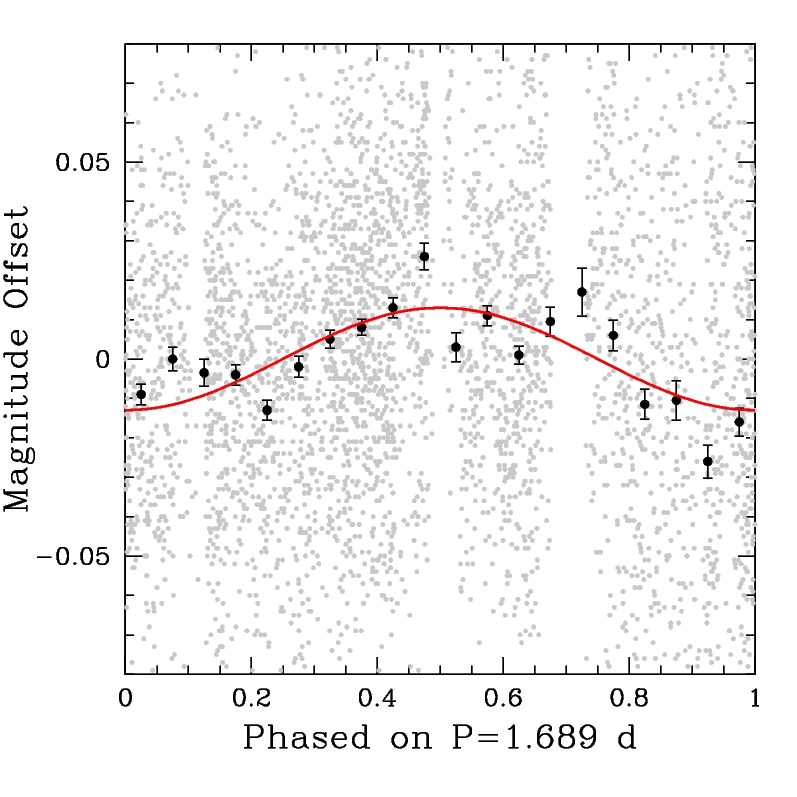}
\caption{\label{stelrot}A preliminary light curve of a star with a candidate 1.689 d rotation period. The light curve has been phased on the shown period. The 4058 epochs are shown in light gray, and the data binned into 20 data points is shown in black. The red line denotes the best fit sine curve to the data with amplitude $0.013$ magnitudes.}
\end{figure}

Rotation is a fundamental feature of stars, and its evolution throughout stellar lifetimes as a function of mass and age is widely studied (see \cite{bouvier14} for a recent review). The primary method for measuring stellar rotation rates is detecting the brightness modulation of stars induced by starspots rotating into and out of the Earth-facing side. The programmatic necessities of these detections, namely high precision and cadence over a baseline of many tens of days, make DEMONEXT a prime instrument for carrying out stellar rotation surveys. As a proof of concept, DEMONEXT has begun systematic monitoring of nearby open clusters with the goal of augmenting the current trove of rotation rates collected in the literature (e.g. \cite{gallet15}).  

We search for rotation periods using a boot-strap Monte-Carlo analysis of the Lomb-Scargle power spectrum \citep{lomb76,scargle82} of each star following the work of \cite{henderson12}. In this analysis, the observation times are not changed but the magnitude values of the light curve are randomly scrambled. The Lomb-Scargle power is re-calculated during each iteration and if the new power is larger than the original power, the candidate period is rejected. In Figure~\ref{stelrot}, we show an example of a detected rotation rate of 1.689 days with amplitude of $0.013$ magnitudes in a possible member of the Alpha Persei cluster. These observations consist of 4,058 individual observations over 12 nights with a 34 day baseline and demonstrate the power of DEMONEXT for uncovering the angular momentum content of nearby stars. 

\section{Conclusions}\label{secconcl}

The Dedicated Monitor of Exotransits and Transients (DEMONEXT) is a 0.5 meter robotic telescope equipped with a dedicated 2k$^2$ CCD imager and 10-position filter wheel. Built to observe transiting exoplanets and transients in an automated and efficient way, DEMONEXT has produced 143 light curves for the KELT collaboration, and 48 light curves for the ASAS-SN supernovae group in the first year of operation. In continuous observing mode, DEMONEXT achieves a raw, unbinned, 2--4\,mmag precision on bright $V<13$ targets with 20--120 second exposures, where most targets are scintillation noise limited. Targets are observed with sub-pixel position stability on the CCD over many hours of continuous observations, and the data can be binned down to 1\,mmag precision on 5--6 minute timescales. For single epoch observations, DEMONEXT achieves 1--10\% relative photometry on most ASAS-SN targets $V<17$ in 5 minute exposures, and is able to detect objects as faint as $V\approx21$. In addition to KELT and ASAS-SN observations, DEMONEXT has also been employed for a number of ancillary science projects, including Galactic microlensing, active galactic nuclei, stellar variability, and stellar rotation, all using the two current observing modes.

\acknowledgements

We are grateful to PlaneWave Instruments for their support with the telescope system hardware and software, and for helping with the alignment and installation of the telescope at Winer Observatory, especially Rick Hedrick and Kevin Ivarson.

We would like to thank Jerry Mason, Tom O’Brien, Dan Pappalardo, Jon Shover, Dave Steinbrecher, and the rest of the staff of the OSU Imaging Sciences Laboratory for technical support during the laboratory integration, testing, packing, shipping and assistance during commissioning.

We would like to thank the OSU Astronomy Department Director of Computing Services David Will for all his help with computer hardware and software issues.

We would like to thank everyone that provided input in targeting, scheduling, and their preliminary results including members of the KELT collaboration: J. Labadie-Bartz, M. Lund, J. Pepper, J. Rodriquez, and D. Stevens; the ASAS-SN team: S. Bose, J. Brown, P. Chen, S. Dong, T. Holoien, C. Kochanek, and K. Stanek; the Open Clusters team: R. Oelkers and G. Sommers; and the microlensing team: A. Gould, M. Penny, J. Yee, and W. Zhu.

Work by S.V.Jr. is supported by the National Science Foundation Graduate Research Fellowship under Grant No. DGE-1343012. Work by B.S.G., as well as a significant fraction of the DEMONEXT hardware costs, were supported by National Science Foundation CAREER Grant AST-1056524. DEMONEXT is also supported from the Vanderbilt Initiative in Data-intensive Astrophysics (VIDA).

\software{Astrometry.net, astropy, CV2, IDL, IDL Astronomy User's Library, MaximDL, numpy, PWI3, PyEphem, Sextractor, STI}

\end{document}